\documentclass[12pt]{iopart}
\usepackage[english]{babel}
\usepackage[utf8]{inputenc}
\usepackage{hyperref}
\usepackage{url}
\usepackage[pdftex]{graphicx}
\usepackage[suffix=]{epstopdf}
\usepackage[T1]{fontenc}
\expandafter\let\csname equation*\endcsname\relax
\expandafter\let\csname endequation*\endcsname\relax
\usepackage{amsmath}
\usepackage[makeroom]{cancel}
\usepackage{dsfont}

\newcommand{\spinup}{|\hspace{-1mm}\uparrow\rangle}
\newcommand{\spindown}{|\hspace{-1mm}\downarrow\rangle}
\newcommand{\half}[1]{\frac{#1}{2}}
\newcommand{\bracket}[3]{\langle #1 | #2 | #3 \rangle}

\makeatletter
\renewcommand\p@figure{Fig. \arabic{figure}\expandafter\@gobble}
\renewcommand\p@equation{(\arabic{equation})\expandafter\@gobble}
\makeatother

\usepackage{color}

\begin{document}
	
	\title[]{A class of states supporting diffusive spin dynamics in the isotropic Heisenberg model}
	
	\author{Marko Ljubotina, Marko Žnidarič and Tomaž Prosen}
	
	\address{Department of Physics, Faculty of Mathematics and Physics, University of Ljubljana, Jadranska 19, 1000 Ljubljana, Slovenia}
	
	\ead{marko.ljubotina@fmf.uni-lj.si}
	
	\begin{abstract}
		
		The spin transport in isotropic Heisenberg model in the sector with zero magnetization is generically super-diffusive. Despite that, we here demonstrate that for a specific set of domain-wall-like initial product states it can instead be diffusive. We theoretically explain the time evolution of such states by showing that in the limiting regime of weak spatial modulation they are approximately product states for very long times, and demonstrate that even in the case of larger spatial modulation the bipartite entanglement entropy grows only logarithmically in time. In the limiting regime we derive a simple closed equation governing the dynamics, which in the continuum limit and for the initial step magnetization profile results in a solution expressed in terms of Fresnel integrals.
		
	\end{abstract}
	
	\vspace{1cm}
	
	
	\submitto{\jpa}
	
	\maketitle
	
	\section{Introduction}
	
	The isotropic Heisenberg chain is a premier theoretical model of quantum many-body physics. It is in principle solvable by the Bethe ansatz~\cite{Bethe}, but in-spite of that, its transport properties are in general still beyond the reach of exact calculations. This in particular holds for magnetization (or spin) transport, for which at half-filling (the sector with zero magnetization) all known (quasi)local conservation laws are orthogonal to the current \cite{ql} and the Mazur bound~\cite{zotos} can not be used to infer the ballistic transport. Likewise, recently introduced hydrodynamic approaches~\cite{doyon,bertini,enej,moore} based on the thermodynamic Bethe ansatz give no prediction at the isotropic point and half-filling.
	
	Numerical studies on the other hand show super-diffusive magnetization transport at high energy, by either studying non-equilibrium steady state situation in terms of boundary driven Lindblad master equation~\cite{prl11}, unitary evolution of initial states~\cite{Gobert,Ours17}, as well as the current autocorrelation functions~\cite{robin12,bojan13}. Super-diffusive transport is also indirectly hinted at by vanishing of  spin Drude weight (in the anisotropic Heisenberg $XXZ$ model when the anisotropy parameter $\Delta\nearrow 1$)~\cite{drude1, drude2, drude3, drude4, drude5, drude6} and at the same time diverging diffusion constant as $\Delta\searrow 1$~\cite{prl11,Ours17} (also seen in Refs. \cite{diver1, diver2, diver3} as an increasing diffusion constant as $\Delta$ decreases towards 1). For a discussion of transport properties of the anisotropic Heisenberg model see e.g. references in Ref.~\cite{affleck11}.
	
	The properties mentioned above hold for generic states (i.e., at high temperature). It is well known that some special states, an extreme example would be the ground state, can on the other hand show very different behaviour. To give a few examples, Lindblad boundary-driven isotropic model at maximal driving can exhibit sub-diffusive transport~\cite{prosen11}, while a domain wall state in the gapped anisotropic model exhibits exponential inhibition of transport~\cite{Gobert,jesenko11}.
	
	However, specific situations in which non-equilibrium dynamics can be treated either exactly or very efficiently are potentially very valuable. In the present paper we shall analytically and numerically show that there is a simple class of initial product states, namely the states which represent tilted domain wall states, for which dynamics in the isotropic Heisenberg model is diffusive and which remain very close to a product state for exponentially long times.
	
	\subsection{The model}
	
	In this work we study the time evolution of a specific family of states in the one dimensional $XXZ$ Heisenberg spin $1/2$ model with $n$ lattice sites,
	\begin{equation}
	H=J\sum_{x=-n/2+1}^{n/2-1}\left(s_{x}^{\rm x}s_{x+1}^{\rm x}+s_{x}^{\rm y}s_{x+1}^{\rm y}+\Delta s_{x}^{\rm z}s_{x+1}^{\rm z}\right)\,.
	\label{eq1}
	\end{equation}
	Here the spin $1/2$ operators are expressed in term of Pauli matrices as $s_k^\gamma=\half{1}\sigma_k^\gamma$, $\gamma\in \{{\rm x,y,z}\}$. The particular states of interest that we shall study are tilted domain wall states, defined as 
	\begin{equation}
	|\psi(t=0)\rangle=\left(\cos\half{\theta_0}\spinup+\sin\half{\theta_0}\spindown\right)^{\otimes\half{n}}\otimes\left(\sin\half{\theta_0}\spinup+\cos\half{\theta_0}\spindown\right)^{\otimes\half{n}}
	\label{eq2}
	\end{equation}
	where $\theta_0=0$ corresponds to the fully polarised domain wall\footnote{We note that we shall later present a more general class of states which can be efficiently and accurately treated.}. Domain wall evolution has been studied in the past for the isotropic model\cite{Gobert,Ours17}, and even more so for the anistropic Heisenberg model~\cite{Antal99,Santos08,Lancaster10,Mossel10,Sabetta13,Halimeh14}. Such a state has different initial magnetisations in the two halves of the chain, $\langle s_{x\ge 0,<0}^{\rm z}\rangle=\pm\half{1}\cos\theta_0$, and therefore has a global imbalance of magnetization and is thus a suitable initial state for the study of magnetization transport. Observe that  the state in each of the half-chains is an eigenstate of the corresponding Hamiltonian $H$ for any $\theta_0$ if $\Delta=1$.
	
	We are going to study the evolution in time of such initial states in the isotropic model $\Delta=1$. We set $J=\hbar=1$ by fixing units. We shall in particular focus on the time-dependent magnetization profiles
	\begin{equation}
	s(x,t)=\bracket{\psi(t)}{s_x^{\rm z}}{\psi(t)},
	\end{equation}
	and the local magnetization current
	\begin{equation}
	j_x=s_{x-1}^{\rm x}s_{x}^{\rm y}-s_{x-1}^{\rm y}s_{x}^{\rm x}\,,\quad j(x,t)=\bracket{\psi(t)}{j_x}{\psi(t)}.
	\label{eq3}
	\end{equation}
	The total magnetisation of the system $M=\sum_xs_x^{\rm z}=M_{\rm L}+M_{\rm R}$ is a conserved quantity $[H,M]=0$, $M_{\rm L}=\sum_{x< 0} s_x^{\rm z}$, and therefore the local magnetization satisfies the continuity equation $ds_x^{\rm z}/dt+j_{x+1}-j_{x}=0$. The transport type can be determined by e.g. studying the asymptotic scaling of the profiles $s(x,t)$, or, equivalently, by looking at the transferred magnetization $M_{\rm L}(t=0)-M_{\rm L}(t)=M_{\rm R}(t)-M_{\rm R}(t=0)$ between the two halves. The expectation value of this operator can be expressed also as an integral of the local current
	
	\begin{equation}
	\Delta s(t)=\int_0^t j(0,t')dt'\propto t^\alpha\, ,
	\label{eq4}
	\end{equation}
	where $j(0,t')$ is the current at half-cut at time $t'$. For $\alpha=1$ the transport is ballistic, for $\half{1}<\alpha<1$ it is super-diffusive, and for $\alpha=\half{1}$ the transport is diffusive. We again remind the reader that in the sector with $M=0$ and at high temperatures (i.e., generic initial states) the magnetization transport in the isotropic Heisenberg model is super-diffusive with $\alpha \approx \frac{2}{3}$~\cite{prl11,Ours17}, and, interestingly, hydrodynamic (large-scale) evolution of spin density profiles is described by a scaled diffusion equation~\cite{Ours17}.
	
	We are first going to numerically demonstrate that for particular initial states, given by Eq.~\ref{eq2}, the transport is diffusive and that the magnetization profiles can be at large times described by a scaling form $s(x,t)=g(x/t^\alpha)$, with $\alpha=\frac{1}{2}$. We shall then give theoretical explanation, relying on a fact that the state $|\psi(t)\rangle$ remains almost a product state even for very large times. Lastly, we shall comment on the applicability of our theoretical description for more generic initial states and for the anisotropic Heisenberg model.
	
	\section{Numerical simulations}
	
\begin{figure}[ht!]
		\centering
		\includegraphics[width=0.49\textwidth]{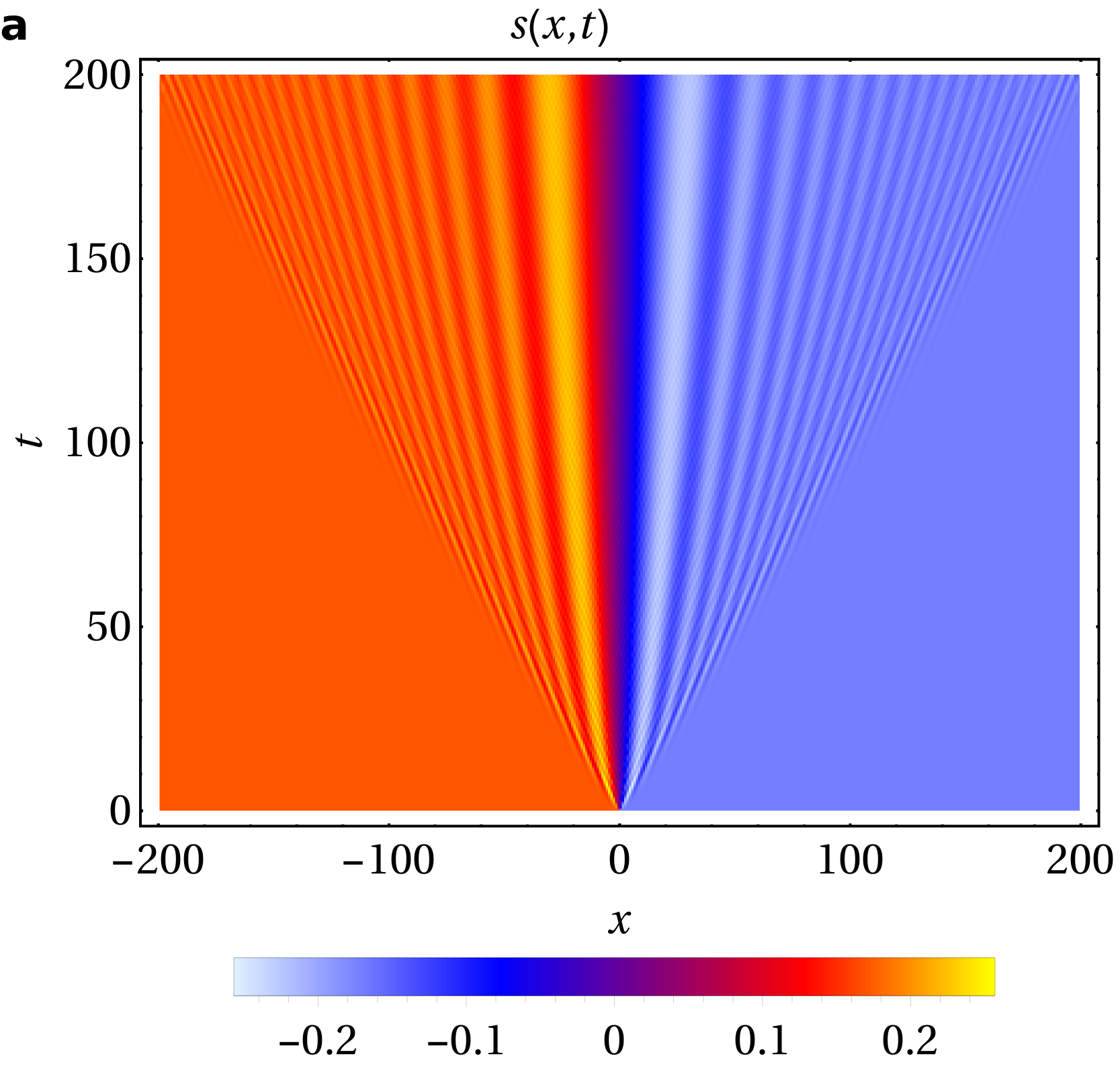}
		\includegraphics[width=0.49\textwidth]{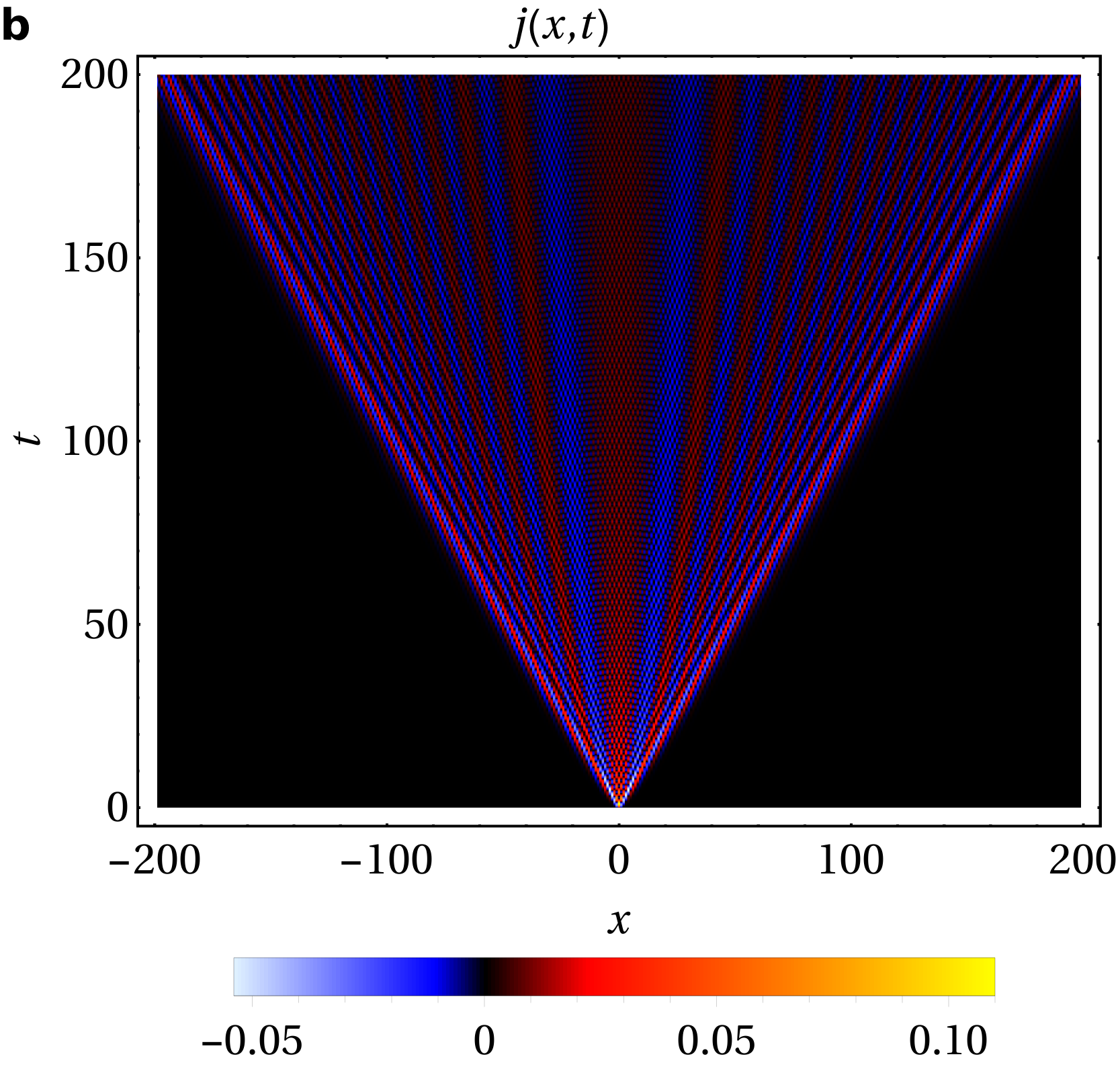}
		\caption{Time evolution of the local magnetization density $s(x,t)$ \textbf{(a)} and current density $j(x,t)$ \textbf{(b)} for an initial state parametrised by $\theta_0=\frac{7\pi}{18}$. Two things can be observed: (i) a diffusive $x \sim t^{1/2}$ spreading of peaks/troughs in the magnetization/current profiles, and (ii) a trivial ballistic light-cone spreading with velocity $v=1$, related to a maximal propagation speed of one spin-flip per unit of time.}
		\label{fig1}
	\end{figure}

\begin{figure}[ht!]
		\centering
		\includegraphics[width=0.49\textwidth]{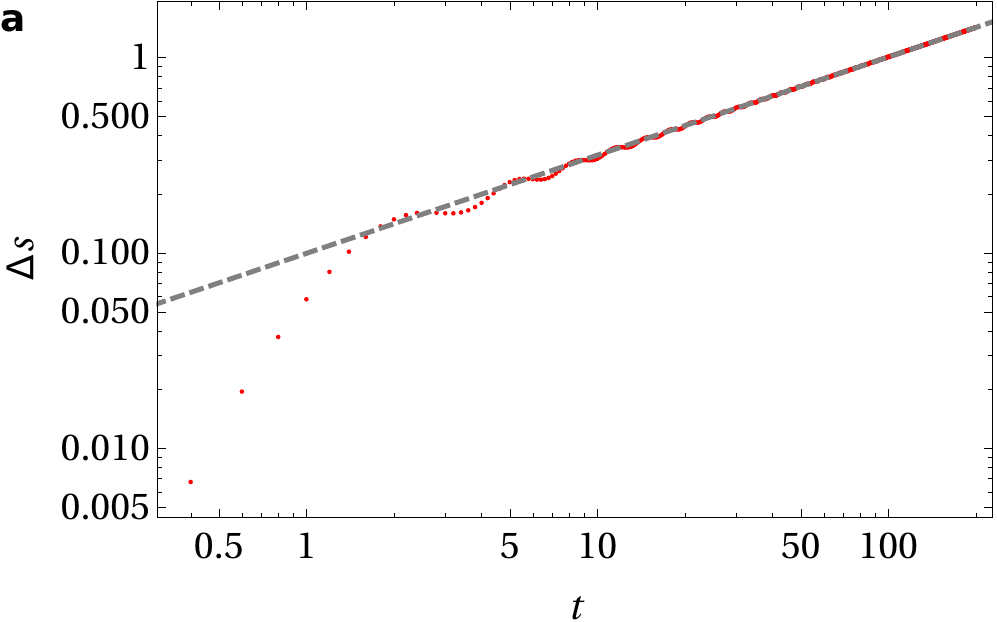}
		\includegraphics[width=0.49\textwidth]{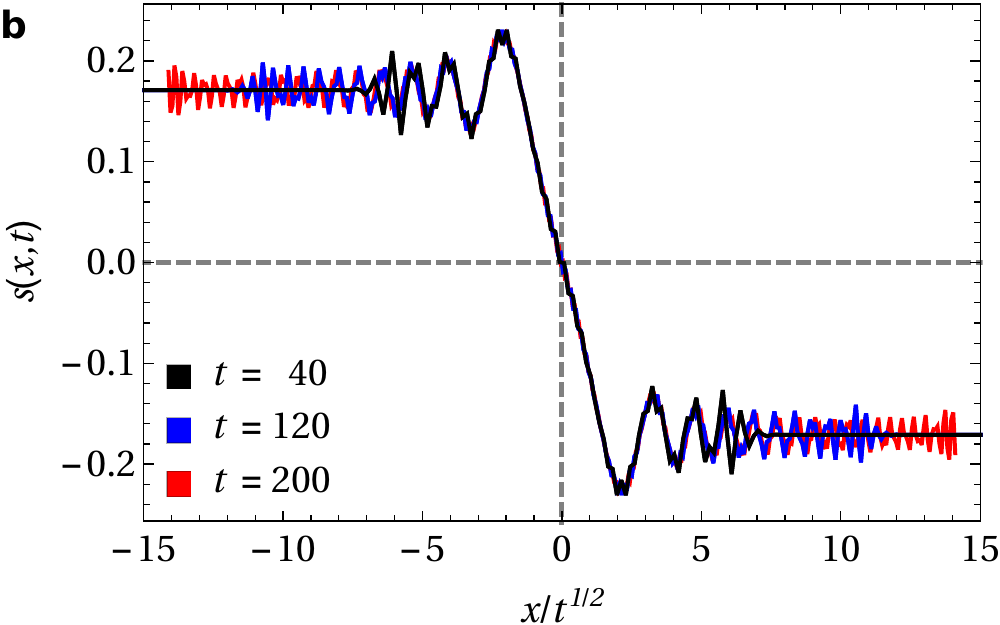}
		\caption{\textbf{(a)} Spin transport through the half-cut, the dashed line shows a power law, Eq.~\ref{eq4}, with the power $\alpha=0.5$. \textbf{(b)} Diffusively scaled spin profiles; the edges, also visible in \ref{fig1} as a ballistic light-cone and here as a wave-front, move to infinity with time. Both figures are for $\theta_0=\frac{7\pi}{18}$ and a system size of $n=400$.}
		\label{fig2}
	\end{figure}

	We study the time evolution $|\psi(t)\rangle = \exp(-{\rm i} H t)|\psi\rangle$  by means of a time-dependent density matrix renormalization group (tDMRG) method, also known as TEBD algorithm \cite{dmrg1, dmrg2, dmrg3}. Choosing an initial state $|\psi(t=0)\rangle$, Eq.~\ref{eq2}, with some particular $\theta_0$ and $\Delta=1$ we show  in \ref{fig1} an example of the time evolution of magnetization and current densities. 
	We see that the edges of the propagation front move ballistically with velocity $v=1$, while more interestingly, the bulk of the magnetization within the light-cone moves considerably slower. This slower transport is in fact diffusive, as can be nicely seen in \ref{fig2}
	where we show scaled profiles at three different times. The choice of the scaled variable $\xi=x/t^{1/2}$ is a clear indication of diffusive transport for such an initial state. 
	
\begin{figure}[ht!]
		\centering
		\includegraphics[width=0.49\textwidth]{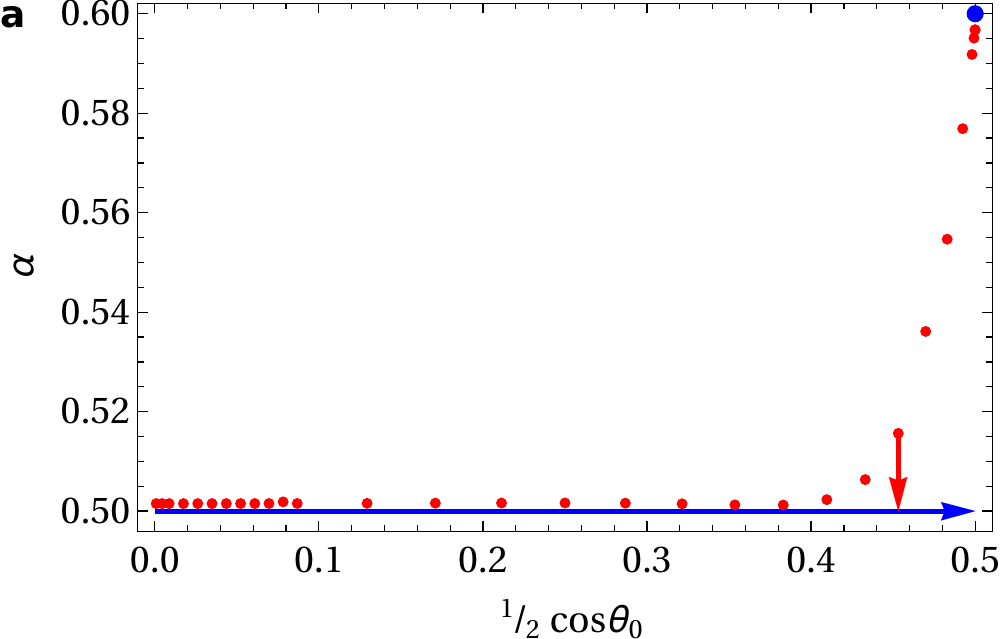}
		\includegraphics[width=0.49\textwidth]{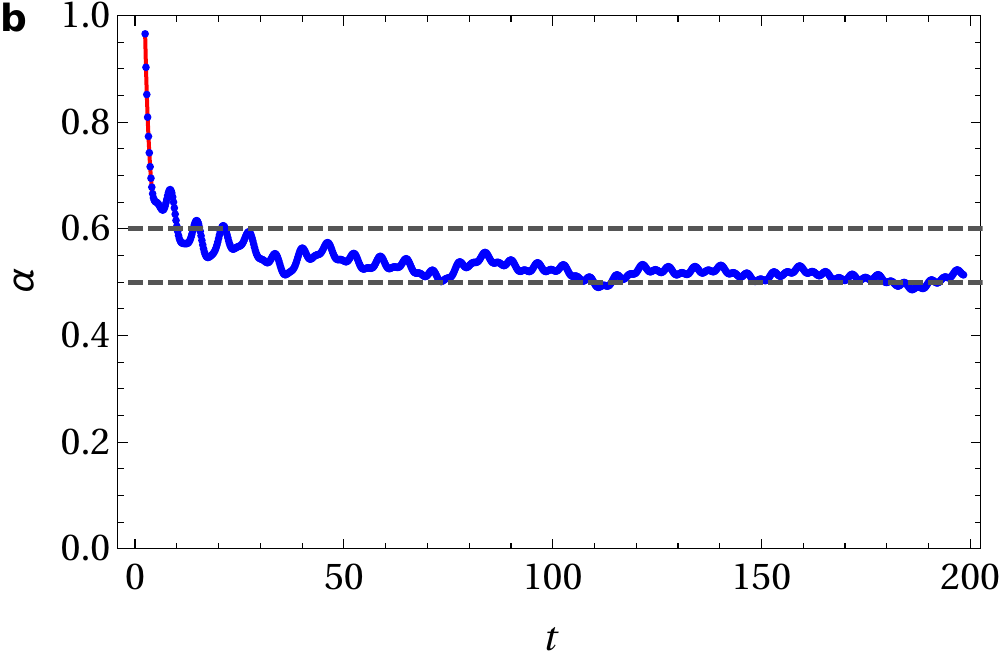}
		\caption{\textbf{(a)} Spin transport exponent $\alpha$ with respect to the initial magnetisation of the two halves. The blue arrow at $\alpha=\frac{1}{2}$ and the point at $\theta_0=0$ denoting $\alpha=\frac{3}{5}$ show our conjecture for the asymptotic $\alpha$ based on detailed analysis of the numerical data (red points). The points near $\frac{1}{2}\cos\theta_0=\frac{1}{2}$ appear to still be converging to the asymptotic $\alpha=0.5$. \textbf{(b)} Convergence of $\alpha(t)$ for the point indicated by the red arrow in frame \textbf{(a)}.}
		\label{fig3}
	\end{figure}

	Repeating the simulation for different values of $\theta_0$ we obtain the exponent $\alpha$ shown in \ref{fig3}\textbf{(a)}. One can see that, except very close to $\theta_0=0$ (fully polarized domain), one always obtains the same diffusive $\alpha=1/2$. Looking at a time-dependent exponent $\alpha(t)$, calculated as a numerical log-derivative $\text{d}\log\Delta s(t)/\text{d}\log t$, \ref{fig3}\textbf{(b)}, we see in fact that even for large $\cos{\theta_0}$ the exponent $\alpha$ also seems to converge to $\frac{1}{2}$ on a long time-scale that can be larger that our maximal simulation time ($t \approx 200$ for small $\theta_0$). Based on that we conjecture that the dynamics is asymptotically diffusive for all $\theta_0 > 0$ ($\alpha=\frac{1}{2}$), while at $\theta_0=0$ (fully polarized domain) it is super-diffusive with exponent $\alpha=\frac{3}{5}$~\cite{Gobert,Ours17}.
	
\begin{figure}[ht!]
		\centering
		\includegraphics[width=0.49\textwidth]{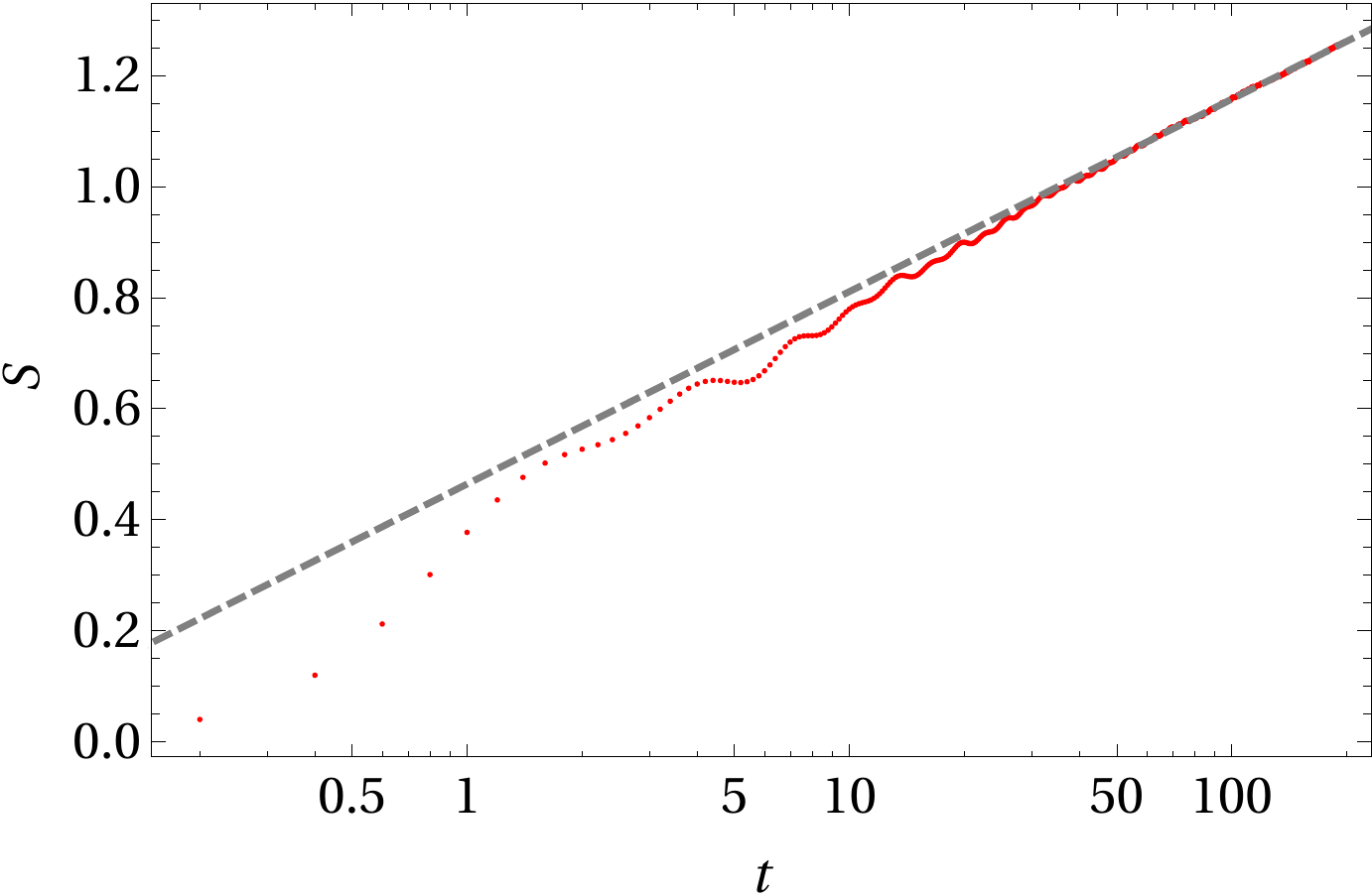}
		\caption{Growth of entanglement entropy $S(t)$ at half-cut for $\theta_0=5\pi/36$. The dashed line indicates $\propto \log t$ growth to guide the eye. }
		\label{fig34}
	\end{figure}

	The reason why we can reach such long simulation times lies in a slow growth of bipartite entanglement entropy, 
	\begin{equation}
	S_{n/2}(t) = -\tr \rho(t)\log \rho(t) = -\sum_{j} \lambda_j^2 \log \lambda_j^2,\quad\rho(t)=
	{\rm tr}_{[-n/2,-1]} |\psi(t)\rangle \langle \psi(t)|,
	\end{equation}
	where $\lambda_j$ are time-dependent Schmidt coefficients obtained as a by-product of the tDMRG algorithm \cite{dmrg1, dmrg2, dmrg3}.
	Our numerical simulations strongly support the conjecture that for any $0 < \theta_0 < \pi$ the entanglement entropy grows logarithmically $S_{n/2}(t) \propto \log t$, or as a very slow power-law (despite very long accessible simulation times it is hard to distinguish the two), see~\ref{fig34}. Note that such growth is typical in local quenches \cite{peschel}. 
	
\begin{figure}[ht!]
		\centering
		\includegraphics[width=0.56\textwidth]{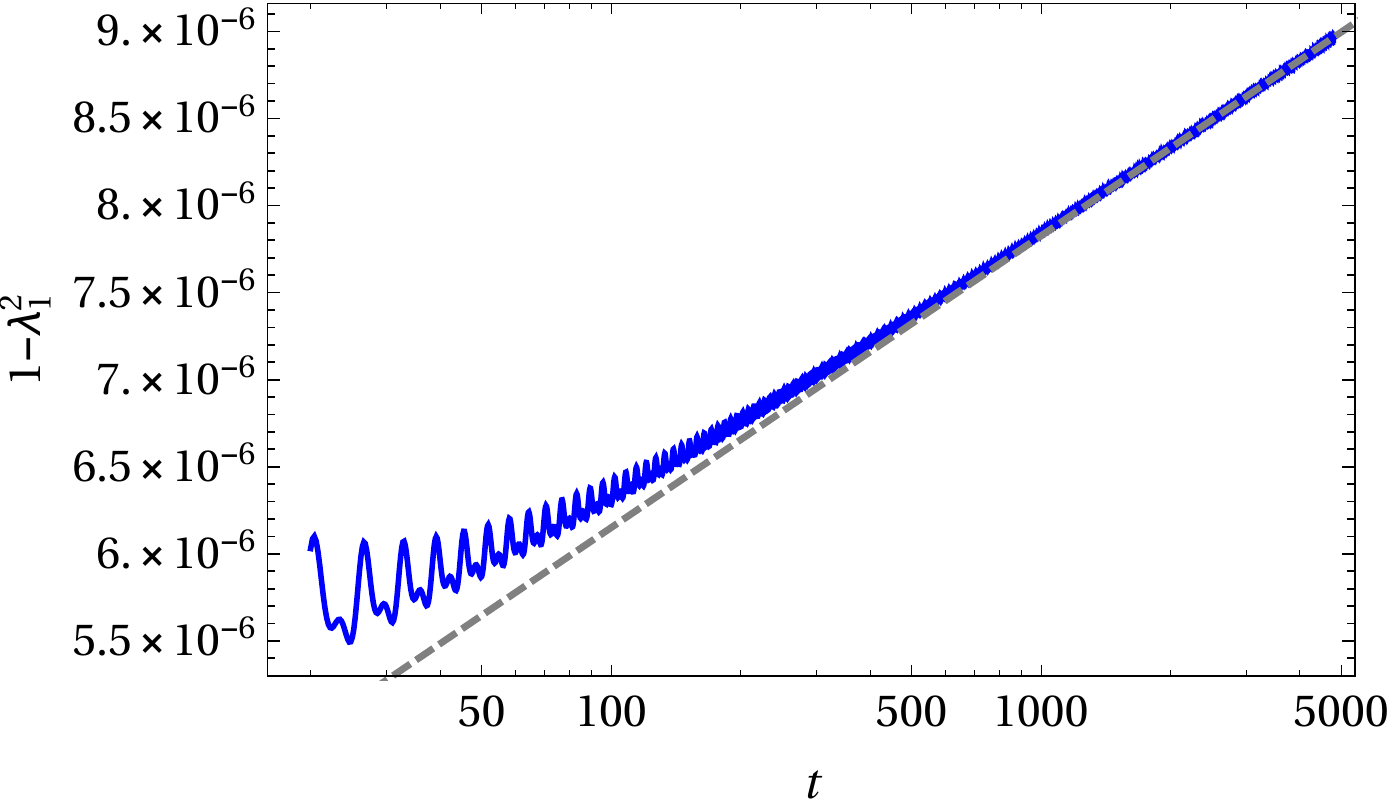}
		\caption{Time evolution of the factorisation error grows extremely slowly. We show data for $\theta_0=\frac{\pi}{2}-\frac{\pi}{36}$. This allows the numerical simulation to easily reach extremely long times. The dashed line indicates $\propto \log t$ growth to guide the eye. }
		\label{fig4}
	\end{figure}

	Another interesting observation is that the state appears to remain approximately a product state up to long times in the limit of small magnetization, $\theta_0\to\frac{\pi}{2}$. \ref{fig4} shows the factorisation error $1-\lambda_1^2$ defined as the sum of squares of Schmidt coefficients
	\begin{equation}
	1-\lambda_1^2=\sum_{k=2}^{M}\lambda_k^2\,,
	\label{eq1v2}
	\end{equation}
	with $M$ being the Schmidt rank. 
	
	In the next section we will explain the observed behaviour in the limiting regime of small $\pi/2-\theta_0$ when the state is almost a product state.
	
	\section{Theoretical explanation}
	
	Observation that the state $\psi(t)$ is almost a product state suggests the following more general product state ansatz
	\begin{equation}
	|\psi_{\rm a}(t)\rangle=\bigotimes_x\left(\cos\frac{\theta_x(t)}{2}\spinup+\sin\frac{\theta_x(t)}{2}e^{i\phi_x(t)}\spindown\right)\,,
	\label{eq5}
	\end{equation}
	with some time-dependent angles $\theta_x(t)$ and $\phi_x(t)$. Note that this is nothing more than a mean-field type ansatz. For the particular class of initial states that we use, Eq.~\ref{eq2}, one has $\phi_x=0$ due to the symmetry. Evaluating the derivative of the wave function for an arbitrary pair of neighbouring sites $(x,x+1)$ we get
	\begin{multline}
	h_{x,x+1}'\left(\cos\frac{\theta_x}{2}\spinup+\sin\frac{\theta_x}{2}\spindown\right)\otimes\left(\cos\frac{\theta_{x+1}}{2}\spinup+\sin\frac{\theta_{x+1}}{2}\spindown\right)=\\
	=\cancel{0|\hspace{-1mm}\uparrow\uparrow\rangle}-\frac{1}{2}\sin\left[\frac{\theta_{x+1}-\theta_{x}}{2}\right]|\hspace{-1mm}\uparrow\downarrow\rangle+\frac{1}{2}\sin\left[\frac{\theta_{x+1}-\theta_{x}}{2}\right]|\hspace{-1mm}\downarrow\uparrow\rangle+\cancel{0|\hspace{-1mm}\downarrow\downarrow\rangle}\,,\qquad\quad\qquad\,\,\,\,
	\label{eq2v2}
	\end{multline}
	where $h_{x,x+1}'=s_x^{\rm x}s_{x+1}^{\rm x}+s_x^{\rm y}s_{x+1}^{\rm y}+s_x^{\rm z}s_{x+1}^{\rm z}-\frac{1}{4}$ is a local 2-site Hamiltonian. A similar type of ansatz \ref{eq5} could be generalized to higher-dimensional Hamiltonians that can be expressed as a sum of local nearest-neighbour permutations, namely, the ordinary Heisenberg Hamiltonian $h'_{x,x+1}$ is an example of such a Hamiltonian that acts on a local Hilbert space of dimensions 2. Writing the Schr\"odinger equation within the ansatz \ref{eq5}, $\frac{{\rm d}}{\rm dt} |\psi_{\rm a}(t)\rangle  = -{\rm i} H |\psi_{\rm a}(t)\rangle$, we can, in the first order approximation in $\epsilon_x=\frac{\pi}{2}-\theta_x$\footnote{Due to rotational invariance we could in fact expand around an arbitrary direction.}, derive the diffusion equations with an imaginary diffusion constant (i.e. single-particle Schr\" odinger equation)
	\begin{equation}
	\frac{\partial \epsilon_x}{\partial t}=-\frac{{\rm i}}{2}(\epsilon_{x-1}-2\epsilon_x+\epsilon_{x+1})\,,
	\label{eq7}
	\end{equation}
	which, in the continuum limit simply reads
	\begin{equation} 
	\partial_t\epsilon=-\frac{{\rm i}}{2}\partial_{xx}\epsilon \,.
	\label{eq7v2}
	\end{equation} 	
	Given our step initial condition \ref{eq2} we can use this differential equation \ref{eq7v2} for $\epsilon$ -- which is, in the limit $\epsilon\to0$ related to the magnetisation as $\langle s_x^z\rangle=\frac{1}{2}\sin\epsilon_x\approx\frac{1}{2}\epsilon_x$ -- to solve for the magnetisation profile
	\begin{equation}
	s(x,t)=\frac{1}{2}\sin\epsilon_0\left[S\left(\frac{x}{\sqrt{\pi t}}\right)+C\left(\frac{x}{\sqrt{\pi t}}\right)\right]\,,
	\label{eq8}
	\end{equation}
	where $\epsilon_0=\frac{\pi}{2}-\theta_0$, and
	$S(z)$ and $C(z)$ are the Fresnel sine and cosine integrals, respectively, defined as 
	\begin{equation}
	C(z)+{\rm i}S(z)=\int_0^ze^{{\rm i}\frac{\pi}{2}\xi^2}d\xi\,.
	\label{eq3v2}
	\end{equation}
	
	\begin{figure}[ht!]
		\centering
		\includegraphics[width=0.49\textwidth]{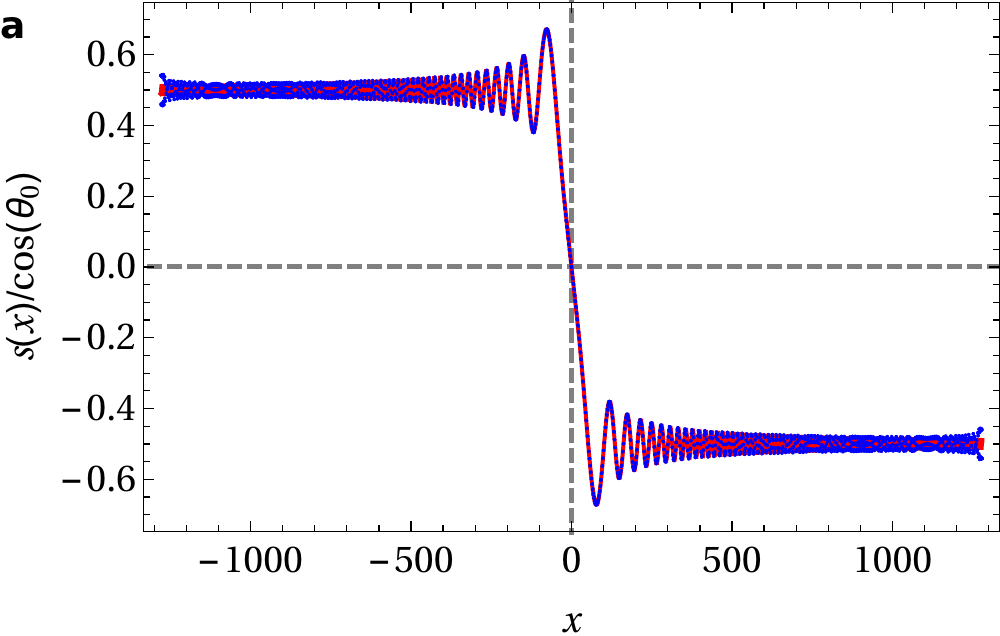}
		\includegraphics[width=0.49\textwidth]{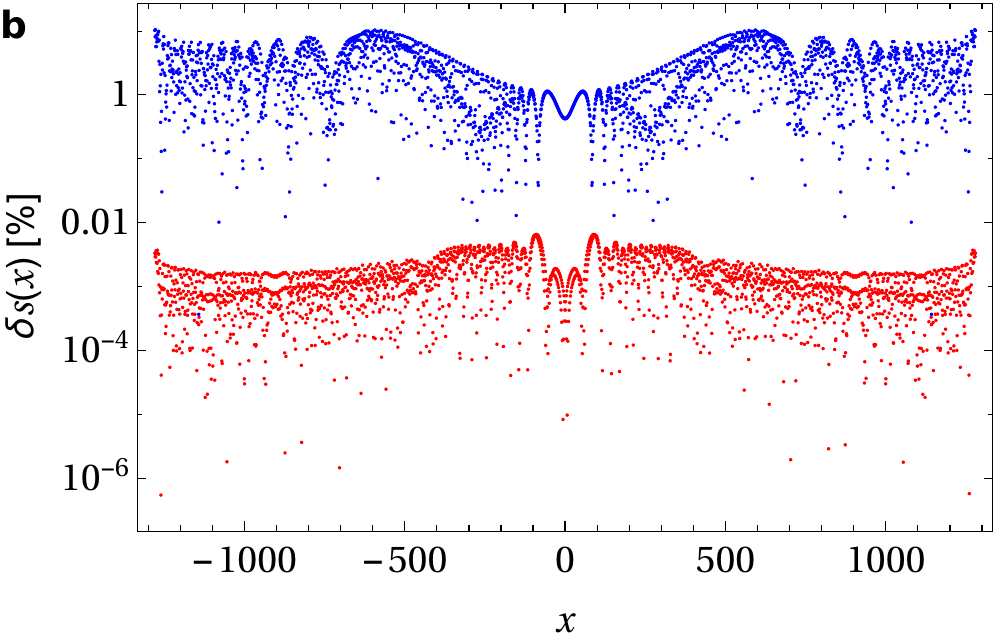}
		\caption{Comparison between the tDMRG results and the theoretical prediction. \textbf{(a)} Spin sprofiles as computed by tDMRG (blue) and theoretical prediction \ref{eq8} (red). \textbf{(b)} Relative error between the two data-sets defined as $\delta s(x)=\frac{2}{\cos\theta_0}|s^{\rm num.}(x)-s^{\rm ana.}(x)|$, the blue points (points with larger errors) represent the continuum limit analytical solution \ref{eq8} and the red points (with relative errors of the order of at most $0.01\%$) are computed numerically for the discrete case \ref{eq7} . 
			We show time $t=1280$ and $\theta_0=\frac{\pi}{2}-\frac{\pi}{180}$.} 
		\label{fig5}
	\end{figure}
	A comparison between the tDMRG results and a numerical solution of equation \ref{eq7} as well as the closed-form solution \ref{eq8} in the continuum limit is shown in \ref{fig5}. One can see that the solution of the difference-differential equation \ref{eq7} describes the exact state almost perfectly, while the continuum solution \ref{eq8} has somewhat larger error, however, the part with larger error is in the tails (e.g., $|x| \ge 500$ in \ref{fig5}) and is pushed towards infinity as time increases. Even though the derivation of equation~\ref{eq8} requires small $\epsilon$, the overall shape of the profile is approximately described by the same theoretical shape given in equation~\ref{eq8} even for larger values of $\epsilon$, see for instance \ref{fig2} where $\theta_0$ is $20^\circ$ above the equator and one still has an oscillatory domain wall shape.

	\section{Stability of the approximate solution and generalizations}
	
	We are now going to discuss the stability of our approximate solution derived in the previous section with respect to varying the initial state as well as the anisotropy $\Delta$. This will give us some information on how special is our class of initial states \ref{eq2}.
	
	The accuracy of the approximate solution with respect to parameters $\gamma_x=\epsilon_{x+1}-\epsilon_x$ (an angle between two consecutive spins in the Bloch representation) and anisotropy $\Delta$ may be studied in the same way by looking at the 2-body Hamiltonian. Performing a small step in time analytically $|\psi(t+\Delta t)\rangle=N(|\psi\rangle+iH\Delta t|\psi(t)\rangle)$, where $N$ is the normalisation and $|\psi\rangle=(\cos(\frac{\pi}{4}+\frac{\gamma_x}{2})\spinup+\sin(\frac{\pi}{4}+\frac{\gamma_x}{2})\spindown)\otimes(\cos(\frac{\pi}{4}-\frac{\gamma_x}{2})\spinup+\sin(\frac{\pi}{4}-\frac{\gamma_x}{2})\spindown)$, we then find the Schmidt decomposition of the resulting state $|\psi(t+\Delta t)\rangle$ where the largest singular value $\lambda_1$ belongs to the product ansatz and the next smaller $\lambda_2$ gives the error we make by assuming a product state ansatz,
	\begin{align}
	\begin{split}
	\frac{\lambda_{2}^2}{\Delta t^2}=&\left[\left(\frac{\Delta-1}{4}\right)^2\right]+\\+&\left[-\left(\frac{\Delta-1}{4}\right)-2\left(\frac{\Delta-1}{4}\right)^2\right]\gamma_x^2+\\+&\left[\frac{4}{3}\left(\frac{\Delta-1}{4}\right)+\frac{5}{3}\left(\frac{\Delta-1}{4}\right)^2\right]\gamma_x^4+\frac{\gamma_x^4}{4}+\mathcal{O}\left(\gamma_x^6\right)\,.
	\end{split}
	\label{eq9}
	\end{align}	
	\begin{figure}[ht!]
		\centering
		\includegraphics[width=0.49\textwidth]{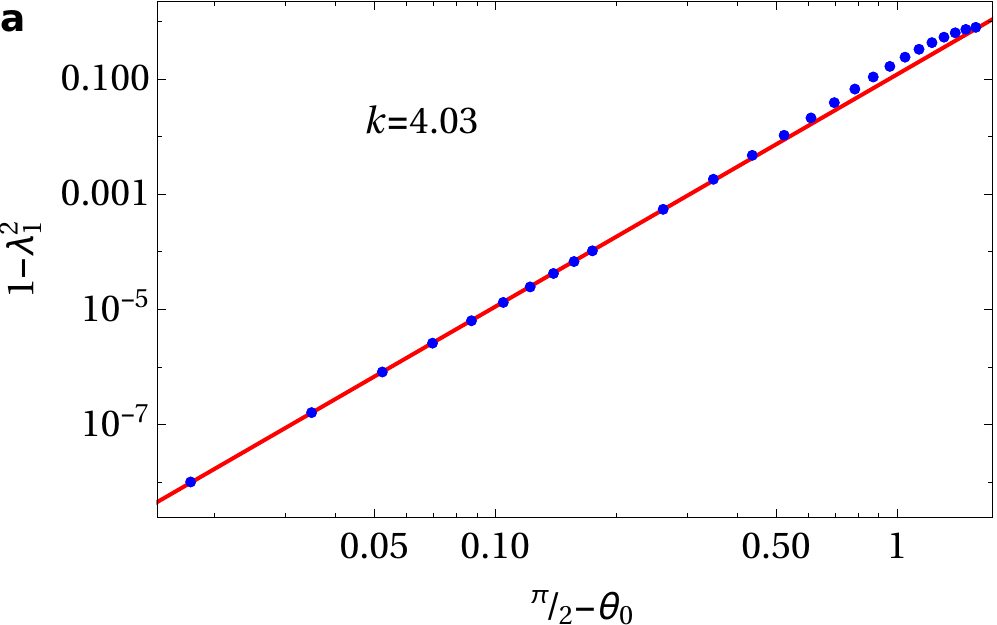}
		\includegraphics[width=0.49\textwidth]{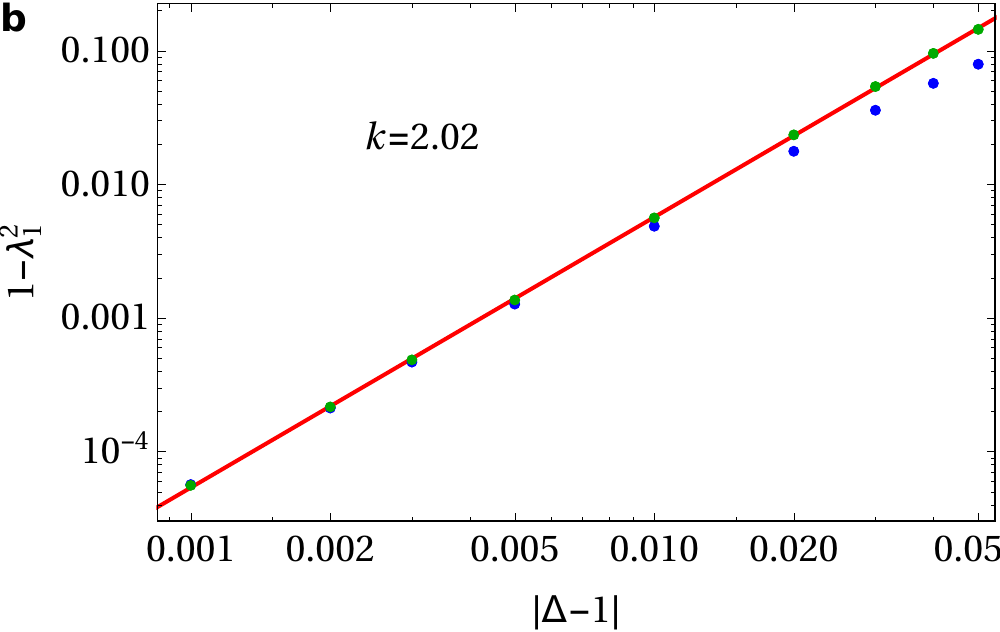}
		\caption{Factorizability of the time evolved state. Size of the factorisation error with respect to initial magnetisation jump $\gamma_0$ (\textbf{(a)}, $t=200$, $\Delta=1$), and anisotropy $\Delta$ (\textbf{(b)}, $t=50$, $\theta_0=\frac{\pi}{2}-\frac{\pi}{36}$), where the blue and green points (matching perfectly with the line) show $\Delta<1$ and $\Delta>1$ respectively. We observe excellent agreement with the theoretical prediction from equation \ref{eq9}, quartic dependence on $(\frac{\pi}{2}-\theta_0)^4$ at the isotropic point, and leading quadratic dependence on anisotropy $|\Delta-1|^2$. The initial state is given by Eq.\ref{eq2}.}
		\label{fig6}
	\end{figure}
	We see that the isotropic case $\Delta=1$ for which we demonstrated diffusive evolution and very slow growth of entanglement is special because the first two orders in $\gamma$ are zero. Using the factorisation error \ref{eq1v2} as a criterion of factorizability of the state, we show in \ref{fig6} its dependence on the initial angle $\theta_0$ as well as on deviations from the isotropic point $|\Delta-1|$. We see that the above equation reproduces the observed power-law errors in the product state ansatz. 
	
	Unfortunately, this approach can not tell us much about the rate at which these errors will accumulate at later times and hence cannot be used to derive the conjectured logarithmic growth of entanglement entropy. In our simple tilted domain wall case, only a single pair of sites has a non-zero value of $\gamma$ in the beginning which represents a sort of {\em local quench} and likely plays an important role in being able to describe such long times both analytically and numerically.
	
	We note that equations \ref{eq7} and \ref{eq9} hold for an arbitrary initial state (not just the one in Eq.\ref{eq2}) if the angles between consecutive spins $|\gamma_x|$ (once again in the Bloch representation of a two-level system) are sufficiently small for all $x$. So one could in principle use \ref{eq7} to evolve a more generic initial product state than ours \ref{eq2} by simply locally rotating the spins to a different basis such that all three relevant $\epsilon_x$ are small \ref{eq7}. However, because the local rotations can be different at different sites the equations are in general non-linear. There is an additional important difference for such an initial product state as compared to \ref{eq2}: the entanglement entropy $S_{n/2}(t)$ will grow initially quadratically in time (instead of log-like as in \ref{fig34}) and the state remains an approximately a product state for considerably shorter times than the state \ref{eq2} (note that for sufficiently small angles this time can still be very large. For example, we can reach $t\approx10^4$ for $\gamma_x=\frac{\pi}{1800}$, where $\gamma_x$ is the angle between two consecutive spins of an initial product state parametrized by a random walk on the surface of the Bloch sphere).  
	
	We suggest that a similar procedure should be possible for the anisotropic $XXZ$ chain. The difference being that the initial state of a half-chain is no longer a spatially homogeneous state, but rather the so-called {\em spin-helix state} \cite{Popkov}, which is an eigenstate of the XXZ Hamiltonian (up to boundary terms), described by Bloch-angles profile
	\begin{equation}
	\left\{\begin{matrix}
	\phi_{x+1}=\phi_x\pm\arccos(\Delta)\quad\theta_{x+1}=\theta_x\quad;\,\Delta<1\\\phi_{x+1}=\phi_x\qquad\theta_{x+1}=\arccos\left(\frac{2\left(\cos(\theta_x)\pm\sqrt{\Delta^2-1} \Delta\sin^2(\theta_x)\right)}{-\left(\Delta^2-1\right)\cos(2\theta_x)+\Delta^2+1}\right)\quad;\,\Delta>1
	\end{matrix}\right.\,.
	\label{eq10}
	\end{equation}
	Note that one should fix the sign in $\pm$ across individual half-chains in order for it to remain stable for long times. The initial state of the entire chain would then be a concatenation of two such chains with a sole (small) defect in $\phi_x$ or $\theta_x$ at the connecting pair of spins. 
	
	\section{Discussion}
	
	We have shown diffusive spin transport behaviour and logarithmic increase of an entanglement entropy for a specific set of initial states in the Heisenberg spin-1/2 chain Hamiltonian which otherwise supports super-diffusion. We have derived an approximate analytical formula describing the time evolution of such states expressing magnetization profiles in terms of Fresnel integrals.
	
	It is an interesting open question if we can explicitly construct other special states which would exhibit other types of transport, say ballistic or super-diffusive. 
	
	The authors would like to  acknowledge support from ERC grant OMNES, as well as grants J1-7279, N1-0025, and P1-0044 of the Slovenian Research Agency.


\section*{References}
	
	\begin{thebibliography}{99}
		
		\bibitem{Bethe} H.~Bethe, On the Theory of Metals, I. Eigenvalues and Eignefunctions of a Linear Chain of Atoms, {\em Zeits. Physik} {\bf 74}, 205-226 (1931).

		\bibitem{ql} E.~Ilievski, M.~Medenjak, T.~Prosen, and L.~Zadnik, Quasilocal charges in integrable lattice systems, {\em J. Stat. Mech.} {\bf 2016}, 064008 (2016).

		\bibitem{zotos} X.~Zotos, F.~Naef, and P.~Prelov\v sek, Transport and conservation laws, {\em Phys. Rev. B} {\bf 55}, 11029-11032 (1997).

		\bibitem{doyon}  O.~A.~Castro-Alvaredo, B.~Doyon and T.~Yoshimura, Emergent hydrodynamics in integrable quantum systems out of equilibrium, {\em Phys. Rev. X} {\bf 6}, 041065 (2016).

		\bibitem{bertini}  B.~Bertini, M.~Collura, J.~De Nardis, M.~Fagotti, Transport in out-of-equilibrium XXZ chains:  exact profiles of charges and currents, {\em Phys. Rev. Lett.} {\bf 117}, 207201 (2016).

		\bibitem{enej} E.~Ilievski and J.~De Nardis, Microscopic Origin of Ideal Conductivity in Integrable Quantum Models, {\em Phys. Rev. Lett.} {\bf 119}, 020602 (2017).

		\bibitem{moore} V. B. Bulchandani, R.~Vasseur, C.~Karrasch and J.~E.~Moore, Bethe-Boltzmann Hydrodynamics and Spin Transport in the XXZ Chain, preprint at {\tt  http://arXiv.org/abs/1702.06146} (2017).

		\bibitem{prl11} M.~\v Znidari\v c, Spin transport in a one-dimensional anisotropic Heisenberg model, {\em Phys. Rev. Lett.} {\bf 106}, 220601 (2011).

		\bibitem{Gobert} D.~Gobert, C.~Kollath, U.~Schollw\" ock, and G.~Sch\" utz, Real-time dynamics in spin-1/2 chains with adaptive time-dependent density matrix renormalization group, {\em Phys. Rev. E} \textbf{71}, 036102 (2005).

		\bibitem{Ours17} M. Ljubotina, M. Žnidarič and T. Prosen, Spin diffusion from an inhomogeneous quench in an integrable system, {\em Nat. Commun.} {\bf 8}, 16117 (2017). 

		\bibitem{robin12} R. Steinigeweg, Spin transport in the XXZ model at high temperatures: Classical dynamics vs. quantum S=1/2 autocorrelations, {\em EPL (Europhysics Letters)} {\bf 97}, 67001 (2012).

		\bibitem{bojan13} T.~Prosen and B.~\v Zunkovi\v c, Macroscopic Diffusive Transport in a Microscopically Integrable Hamiltonian System, {\em Phys.~Rev.~Lett.} {\bf 111}, 040602 (2013).

		\bibitem{drude1} T.~Prosen and E.~Ilievski, Families of Quasilocal Conservation Laws and Quantum Spin Transport, {\em Phys.~Rev.~Lett.} {\bf 111}, 057203 (2013).
		
		\bibitem{drude2} R.~Steinigeweg, J.~Gemmer, and W.~Brenig, Spin and energy currents in integrable and nonintegrable spin-$\frac{1}{2}$ chains: A typicality approach to real-time autocorrelations, {\em Phys.~Rev.~B} {\bf 91}, 104404 (2015).
		
		\bibitem{drude3} J.~M.~P.~Carmelo, T.~Prosen, and D.~K.~Campbell, Vanishing spin stiffness in the spin-$\frac{1}{2}$ Heisenberg chain for any nonzero temperature, {\em Phys.~Rev.~B} {\bf 92}, 165133 (2015).
		
		\bibitem{drude4} C.~Karrasch, J.~Hauschild, S.~Langer, and F.~Heidrich-Meisner, The Drude weight of the spin-1/2 XXZ chain: density matrix renormalization group versus exact diagonalization, {\em Phys.~Rev.~B} {\bf 87}, 245128 (2013).
		
		\bibitem{drude5} J.~Herbrych, P.~Prelov\v sek, and X.~Zotos, Finite-temperature Drude weight within the anisotropic Heisenberg chain, {\em Phys.~Rev.~B} {\bf 84}, 155125 (2011).
		
		\bibitem{drude6} J.~Herbrych, R.~Steinigeweg, and P.~Prelov\v sek, Spin hydrodynamics in the $S=\frac{1}{2}$ anisotropic Heisenberg chain, {\em Phys.~Rev.~B} {\bf 86}, 115106 (2012).

		\bibitem{diver1} C.~Karrasch, J.~E.~Moore, and F.~Heidrich-Meisner, Real-time and real-space spin and energy dynamics in one-dimensional spin-$\frac{1}{2}$ systems induced by local quantum quenches at finite temperatures, {\em Phys.~Rev.~B} {\bf 89}, 075139 (2014).
		
		\bibitem{diver2} R.~Steinigeweg and J.~Gemmer, Density dynamics in translationally invariant spin-$\frac{1}{2}$ chains at high temperatures: A current-autocorrelation approach to finite time and length scales, {\em Phys.~Rev.~B} {\bf 80}, 184402 (2009).
		
		\bibitem{diver3} R.~Steinigeweg and W.~Brenig, Spin Transport in the $XXZ$ Chain at Finite Temperature and Momentum, {\em Phys.~Rev.~Lett.} {\bf 107}, 250602 (2011).

		\bibitem{affleck11} J.~Sirker, R.~G.~Pereira, and I.~Affleck, Conservation laws, integrability, and transport in one-dimensional quantum systems, {\em Phys. Rev. B} {\bf 83}, 035115 (2011). 

		\bibitem{prosen11} T.~Prosen, Exact Nonequilibrium Steady State of a Strongly Driven Open XXZ Chain, {\em Phys.~Rev.~Lett.} {\bf 107}, 137201 (2011).

		\bibitem{jesenko11} S.~Jesenko and M.~\v Znidari\v c, Finite-temperature magnetization transport of the one-dimensional anisotropic Heisenberg model, {\em Phys.~Rev.~B} {\bf 84}, 174438 (2011).

		\bibitem{Antal99} T.~Antal, Z.~R\' acz, A.~R\' akos, and G.~M.~Sch\" utz, Transport in the XX chain at zero temperature: Emergence of flat magnetization profiles, {\em Phys.~Rev.~E} {\bf 59}, 4912 (1999).

		\bibitem{Santos08} L.~F.~Santos, Transport control in low-dimensional spin-1/2 Heisenberg systems, {\em Phys.~Rev.~E} {\bf 78}, 031125 (2008).

		\bibitem{Lancaster10} J.~Lancaster and A.~Mitra, Quantum quenches in an XXZ spin chain from a spatially inhomogeneous initial state, {\em Phys.~Rev.~E} {\bf 81}, 061134 (2010).

		\bibitem{Mossel10} J.~Mossel and J.-S.~Caux, Relaxation dynamics in the gapped XXZ spin-1/2 chain, {\em New J.~Phys.} {\bf 12}, 055028 (2010).

		\bibitem{Sabetta13} T.~Sabetta and G.~Misguich, Nonequilibrium steady states in the quantum XXZ spin chain, {\em Phys.~Rev.~B} {\bf 88}, 245114 (2013).

		\bibitem{Halimeh14} J.~C.~Halimeh, A.~ W\" ollert, I.~McCulloch, U.~Schollw\" ock, and T.~Barthel, Domain-wall melting in ultracold-boson systems with hole and spin-flip defects, {\em Phys.~Rev.~A} {\bf 89}, 063603 (2014).

		\bibitem{dmrg1} S.~R.~White, Density matrix formulation for quantum renormalization groups, \emph{Phys. Rev. Lett.} \textbf{69}, 2863 (1992).
		
		\bibitem{dmrg2} G.~Vidal, Efficient Classical Simulation of Slightly Entangled Quantum Computations, \emph{Phys. Rev. Lett.} \textbf{91}, 147902 (2003).
		
		\bibitem{dmrg3} U.~Schollw\"ock, The density-matrix renormalization group in the age of matrix product states, \emph{Ann. Phys.} \textbf{326}, 96-192 (2011).
		
		\bibitem{peschel} V.~Eisler and I.~Peschel, Evolution of entanglement after a local quench, \emph{J. Stat. Mech.} {\bf 2007}, P06005 (2007). 

		\bibitem{Popkov} V. Popkov and G. M. Sch\"utz, Solution of the Lindblad equation for spin helix states, \emph{Phys. Rev. E} \textbf{95}, 042128 (2017). 

	\end{thebibliography}
\end{document}